\documentclass[12pt]{article}
\usepackage{graphicx}

\def\nn{\nonumber}
\def\sss{\scriptscriptstyle}
\def\roughly#1{\mathrel{\raise.3ex\hbox
{$#1$\kern-.75em\lower1ex\hbox{$\sim$}}}}

\def\sss{\scriptscriptstyle}
\def\barp{{\raise.35ex\hbox{${\sss (}$}}---{\raise.35ex\hbox{${\sss )}$}}}
\def\bdbarp{\hbox{$B_d$\kern-1.4em\raise1.4ex\hbox{\barp}}}
\def\bsbarp{\hbox{$B_s$\kern-1.4em\raise1.4ex\hbox{\barp}}}
\def\dbarp{\hbox{$D$\kern-1.1em\raise1.4ex\hbox{\barp}}}

\newcommand{\bd}{B_d^0}
\newcommand{\bdb}{\overline{B_d^0}}
\newcommand{\bs}{B_s^0}
\newcommand{\bsbar}{\overline{B_s^0}}

\newcommand{\beq}{\begin{equation}}
\newcommand{\eeq}{\end{equation}}

\newcommand{\abseps}{\vert\epsilon\vert}

\newcommand{\fbb}{f^2_{B_d}\hat{B}_{B_d}}
\newcommand{\fbbs}{f^2_{B_s}\hat{B}_{B_s}}
\newcommand{\fbd}{f_{B_d}}

\def\rly#1{\mathrel{\raise.3ex\hbox{$#1$\kern-.75em\lower1ex\hbox{$\sim$}}}}

\newread\epsffilein 
\newif\ifepsffileok 
\newif\ifepsfbbfound 
\newif\ifepsfverbose 
\newdimen\epsfxsize 
\newdimen\epsfysize 
\newdimen\epsftsize 
\newdimen\epsfrsize 
\newdimen\epsftmp 
\newdimen\pspoints 
\def\epsfbox#1{\global\def\epsfllx{72}\global\def\epsflly{72}%
 \global\def\epsfurx{540}\global\def\epsfury{720}%
 \def\lbracket{[}\def\testit{#1}\ifx\testit\lbracket
 \let\next=\epsfgetlitbb\else\let\next=\epsfnormal\fi\next{#1}}%
\def\epsfgetlitbb#1#2 #3 #4 #5]#6{\epsfgrab #2 #3 #4 #5 .\\%
 \epsfsetgraph{#6}}%
\def\epsfnormal#1{\epsfgetbb{#1}\epsfsetgraph{#1}}%
\def\epsfgetbb#1{%
\openin\epsffilein=#1
\ifeof\epsffilein\errmessage{I couldn't open #1, will ignore it}\else
 {\epsffileoktrue \chardef\other=12
 \def\do##1{\catcode`##1=\other}\dospecials \catcode`\ =10
 \loop
 \read\epsffilein to \epsffileline
 \ifeof\epsffilein\epsffileokfalse\else
 \expandafter\epsfaux\epsffileline:. \\%
 \fi
 \ifepsffileok\repeat
 \ifepsfbbfound\else
 \ifepsfverbose\message{No bounding box comment in #1; using defaults}\fi\fi
 }\closein\epsffilein\fi}%
\def\epsfclipstring{}
\def\epsfsetgraph#1{%
 \epsfrsize=\epsfury\pspoints
 \advance\epsfrsize by-\epsflly\pspoints
 \epsftsize=\epsfurx\pspoints
 \advance\epsftsize by-\epsfllx\pspoints
 \epsfxsize\epsfsize\epsftsize\epsfrsize
 \ifnum\epsfxsize=0 \ifnum\epsfysize=0
 \epsfxsize=\epsftsize \epsfysize=\epsfrsize
 \epsfrsize=0pt
 \else\epsftmp=\epsftsize \divide\epsftmp\epsfrsize
 \epsfxsize=\epsfysize \multiply\epsfxsize\epsftmp
 \multiply\epsftmp\epsfrsize \advance\epsftsize-\epsftmp
 \epsftmp=\epsfysize
 \loop \advance\epsftsize\epsftsize \divide\epsftmp 2
 \ifnum\epsftmp>0
 \ifnum\epsftsize<\epsfrsize\else
 \advance\epsftsize-\epsfrsize \advance\epsfxsize\epsftmp \fi
 \repeat
 \epsfrsize=0pt
 \fi
 \else \ifnum\epsfysize=0
 \epsftmp=\epsfrsize \divide\epsftmp\epsftsize
 \epsfysize=\epsfxsize \multiply\epsfysize\epsftmp
 \multiply\epsftmp\epsftsize \advance\epsfrsize-\epsftmp
 \epsftmp=\epsfxsize
 \loop \advance\epsfrsize\epsfrsize \divide\epsftmp 2
 \ifnum\epsftmp>0
 \ifnum\epsfrsize<\epsftsize\else
 \advance\epsfrsize-\epsftsize \advance\epsfysize\epsftmp \fi
 \repeat
 \epsfrsize=0pt
 \else
 \epsfrsize=\epsfysize
 \fi
 \fi
 \ifepsfverbose\message{#1: width=\the\epsfxsize, height=\the\epsfysize}\fi
 \epsftmp=10\epsfxsize \divide\epsftmp\pspoints
 \vbox to\epsfysize{\vfil\hbox to\epsfxsize{%
 \ifnum\epsfrsize=0\relax
 \includegraphics{#1}%
 \else
 \epsfrsize=10\epsfysize \divide\epsfrsize\pspoints
 \includegraphics{#1}%
 \fi
 \hfil}}%
\global\epsfxsize=0pt\global\epsfysize=0pt}%
 {\catcode`\%=12 \global\let\epsfpercent=
\long\def\epsfaux#1#2:#3\\{\ifx#1\epsfpercent
 \def\testit{#2}\ifx\testit\epsfbblit
 \epsfgrab #3 . . . \\%
 \epsffileokfalse
 \global\epsfbbfoundtrue
 \fi\else\ifx#1\par\else\epsffileokfalse\fi\fi}%
\def\epsfempty{}%
\def\epsfgrab #1 #2 #3 #4 #5\\{%
\global\def\epsfllx{#1}\ifx\epsfllx\epsfempty
 \epsfgrab #2 #3 #4 #5 .\\\else
 \global\def\epsflly{#2}%
 \global\def\epsfurx{#3}\global\def\epsfury{#4}\fi}%
\def\epsfsize#1#2{\epsfxsize}
\def\mt{m_t}

\def\mc{m_c}

\newcommand{\delmd}{\Delta M_d}
\newcommand{\delms}{\Delta M_s}

\pagestyle{plain}

\begin{document}

\begin{flushright}  
DESY 00-182 \\
UdeM-GPP-TH-00-80\\
December 2000\\
\end{flushright}

\begin{center} 

{\large \bf
\centerline{What if the Mass Difference $\Delta M_s$ is around 18
Inverse Picoseconds?}}
\vspace*{1.5cm}
{\large A.~Ali} \vskip0.2cm
Deutsches Elektronen Synchrotron DESY, Hamburg \\
\vspace*{0.3cm} \centerline{ and} \vspace*{0.3cm} {\large D.~London}
\vskip0.2cm Laboratoire Ren\'e J.-A. L\'evesque, Universit\'e de
Montr\'eal, \\
C.P. 6128, succ.\ centre-ville, Montr\'eal, QC, Canada H3C 3J7 \\
\vskip0.5cm
\vskip0.5cm
{\Large Abstract\\}
\vskip3truemm

\parbox[t]{\textwidth} {Present experiments in pursuit of the mass
  difference in the $B_s^0$-$\overline{B_s^0}$ system have put a lower
  bound on this quantity of $\Delta M_s > 14.9 ~{\rm ps}^{-1}$ (at
  95\% C.L.). The same experiments also yield a local minimum in the
  log-likelihood function around $\Delta M_s = 17.7 ~{\rm ps}^{-1}$,
  which is $2.5\sigma$ away from being zero. Motivated by these
  observations, we investigate the consequences of a possible
  measurement of $\Delta M_s = 17.7 \pm 1.4~{\rm ps}^{-1}$, in the
  context of both the standard model and supersymmetric models with
  minimal flavor violation. We perform a fit of the quark mixing
  parameters in these theories and estimate the expected ranges of the
  CP asymmetries in $B$ decays, characterized by $\alpha$, $\beta$ and
  $\gamma$, the interior angles of the CKM-unitarity triangle. Based
  on this study, we argue that, if indeed $\Delta M_s$ turns out to be
  in its currently-favored range, this would disfavor a large class of
  supersymmetric models. Indeed, of all the models examined here, the
  best fit to the data occurs for the standard model.}

\end{center}
\thispagestyle{empty}
\newpage
\setcounter{page}{1}
\textheight 23.0 true cm

%
\baselineskip=14pt
%

\section{Introduction}

One of the principal aims of flavor physics is to measure the
parameters of the Cabibbo-Kobayashi-Maskawa (CKM) matrix \cite{CKM},
which encodes the manner in which quark mixing takes place within the
Standard Model (SM). There are many measurements which contribute to
this goal. For example, the matrix element $V_{ud}$ can be probed
through the study of neutron $\beta$ decay, while the $\bd$--$\bdb$
mass difference $\delmd$ can be used to determine the matrix element
$V_{td}$. Our present knowledge of the CKM matrix is usually displayed
in terms of the allowed region of the so-called unitarity triangle
\cite{PDG00}. Ongoing experiments studying $B$-hadron physics will be
able to test the CKM matrix by measuring the sides and the
(CP-violating) angles of the unitarity triangle. If physics beyond the
SM is present, inconsistencies in the various unitarity tests will
appear. If this occurs, then it will be necessary to perform an
overall fit of the CKM matrix elements in various competing theories
in order to establish the right framework for flavor physics.

One appealing candidate theory which may induce such ``unitarity
inconsistencies'' is supersymmetry (SUSY). In its minimal
flavor-violating form, the couplings of SUSY particles to ordinary
matter are proportional to CKM matrix elements. Thus, the weak phases
of supersymmetric contributions to loop-induced transitions are the
same as in the SM. These loop-level processes include $\bd$--$\bdb$
and $\bs$--$\bsbar$ mixing, as well as the flavor-changing
neutral-current decays $b \to s \gamma$ and $b \to s \ell^+ \ell^-$.
The presence of such additional SUSY contributions has the effect that
the extracted values of the matrix elements $\vert V_{td}\vert$ and
$\vert V_{ts}\vert$ will be modified from their SM values. Conversely,
precise measurements of the CKM matrix elements may put severe bounds
on new physics, including SUSY. In Ref.~\cite{AL99}, we demonstrated
this quantitatively: we worked out the profile of the CKM unitarity
triangle in the SM and in several variants of minimal flavor-violating
supersymmetric models. We also examined the correlations among the
CP-violating phases $\alpha$, $\beta$ and $\gamma$ in these models.
Although, at the present time, all models give reasonable fits to the
data, in the future, with more precise data, one will be able to
distinguish among the various candidate models.

If one compares the allowed region of the unitarity triangle of today
with that of the early 1990's \cite{oldAL} it is clear that the
current region is considerably smaller. Although the errors on
virtually all measurements have decreased since the early 1990's, the
single most important improvement has been the measurement of $\Delta
M_s$ in $\bs$--$\bsbar$ mixing. As the lower limit on $\Delta M_s$ has
increased over the years, more and more of the earlier-allowed region
has been cut away. Indeed, this lower limit continues to increase:
although the lower limit in 1999 was $\delms > 12.4~{\rm ps}^{-1}$, it
now stands at $\delms > 14.9 ~{\rm ps}^{-1}$ \cite{Stocchi00}. More
intriguing, there is now a hint of a possible signal at $\delms \simeq
17.7~{\rm ps}^{-1}$ \cite{Stocchi00}. Clearly, the last word on
$\delms$ from the combined LEP/SLD/CDF analysis is yet to come, and it
is conceivable that the measurement of $\delms$ is just around the
corner. In anticipation of this, and to underscore the importance of
the $\delms$ measurement for CKM phenomenology, in this paper we
present two analyses. First, we update the CKM fits in the SM and in
the supersymmetric models mentioned above. Second, we assume a
(future) measurement of $\Delta M_s = 17.7 \pm 1.4~{\rm ps}^{-1}$, and
examine the consequences. As we will see, such a measurement would be
sufficient to disfavor a large class of minimal flavor-violating
supersymmetric models (though it would be completely consistent with
the SM).

The paper is organized as follows. In the next section, we present the
year-2000 profile of the unitarity triangle, both in the SM and in
supersymmetric theories with minimal flavor violation. As we will see,
the measurement of $\beta$ will not distinguish among these various
models, though the measurement of $\alpha$ and/or $\gamma$ will. More
to the point, in supersymmetric models one can obtain a very different
allowed range for $\Delta M_s$, so that a precision measurement of
this quantity will be able to strongly constrain the SUSY parameter
space. This is shown quantitatively in Sec. 3. Here we present a
future profile of the unitarity triangle, both in the SM and in SUSY,
assuming a hypothetical measurement of $\Delta M_s = 17.7 \pm
1.4~{\rm ps}^{-1}$. Such a measurement would disfavor a certain class
of SUSY models.  Furthermore, it turns out that, of all the models
considered here, the SM yields the best fit to the data. Thus, the
hint of a signal at $\delms = 17.7~{\rm ps}^{-1}$ is not in any way in
conflict with the SM. (Of course, there is still a large class of SUSY
models which provides a reasonable fit to the data.) We conclude in
Sec.~4.

\section{Unitarity Triangle: Year-2000 Profile}

It is customary to use an approximate parametrization of the CKM
matrix, due to Wolfenstein \cite{Wolfenstein}, to quantitatively
discuss the allowed region of the unitarity triangle. The Wolfenstein
parametrization can be written as
\beq
V \simeq \left(\matrix{
 1-{1\over 2}\lambda^2 & \lambda
 & A\lambda^3 \left( \rho - i\eta \right) \cr
 -\lambda ( 1 + i A^2 \lambda^4 \eta )
& 1-{1\over 2}\lambda^2 & A\lambda^2 \cr
 A\lambda^3\left(1 - \rho - i \eta\right) & -A\lambda^2 & 1 \cr}\right)~.
\label{CKM}
\eeq
Thus, $\lambda$, $A$, $\rho$ and $\eta$ are the four quantities which
parametrize the CKM matrix.

With the experimental precision expected in future $B$ (and $K$)
decays, it may become necessary to go beyond leading order in
$\lambda$ in the Wolfenstein parametrization given above. To this end,
we follow here the prescription of Buras et al.\ \cite{BLO94}:
defining $\bar{\rho} \equiv \rho(1-\lambda^2/2)$ and $\bar{\eta}
\equiv \eta(1-\lambda^2/2)$, we have
\beq
V_{us} = \lambda,~~~V_{cb}=A\lambda^2,~~~V_{ub}=A\lambda^3(\rho -i
\eta),~~~V_{td} = A\lambda^3(1- \bar{\rho} -i \bar{\eta})
\label{nlo-wolf}
\eeq
(Note that the matrix elements $V_{us}, V_{cb}$ and $V_{ub}$ remain
unchanged, but $V_{td}$ is renormalized in going from leading order to
next-to-leading order.) The apex of the unitarity triangle is now
defined by the renormalized Wolfenstein parameters
$(\bar{\rho},\bar{\eta})$.

\subsection{Input Data}

There are a variety of measurements which constrain $\bar{\rho}$ and
$\bar{\eta}$, either directly or indirectly. The theoretical and
experimental quantities which are used in the CKM fits are listed in
Table~\ref{datatable}, along with their present values and errors (if
applicable). For a detailed description of these quantities, as well
as a discussion of our methodology, we refer the reader to
Ref.~\cite{AL99}.

\begin{table}[t]
\centering
\caption{ \it Data used in the CKM fits}
\vskip 0.1 in
\begin{tabular}{|c|c|} \hline
Parameter & Value \\
\hline
\hline
$\lambda$ & $0.2196$  \cr
$\vert V_{cb} \vert $ & $0.0404 \pm 0.0018$ \cr
$\vert V_{ub} / V_{cb} \vert$ & $0.087 \pm 0.018$ \cr
$\abseps$ & $(2.280 \pm 0.013) \times 10^{-3}$ \cr
$\Delta M_d$ & $0.487 \pm 0.014~{\rm ps}^{-1}$ \cr
$\Delta M_s$ & $ > 14.9 ~{\rm ps}^{-1}$ \cr 
$\overline{\mt}(\mt(pole))$ & $165 \pm 5$ GeV  \cr
$\overline{\mc}(\mc(pole))$ & $1.25 \pm 0.05$ GeV  \cr
$\hat{\eta}_B$ & $0.55$ \cr
$\hat{\eta}_{cc} $ & $1.38 \pm 0.53$  \cr
$\hat{\eta}_{ct} $ & $0.47 \pm 0.04$  \cr
$\hat{\eta}_{tt} $ & $0.57$  \cr
$\hat{B}_K$ & $0.94 \pm 0.15$  \cr
$\fbd\sqrt{\hat{B}_{B_d}} $ & $230 \pm 40$ MeV  \cr
$\xi_s $ & $1.16 \pm 0.05$   \cr
\hline
\end{tabular}
\label{datatable}
\end{table}

The one measurement which must be described in more detail here is
$\Delta M_s$. Since the first studies of $\bs$-$\bsbar$ mixing in the
SM \cite{ali-aydin}, it was known that the measurement of the mass
differences $\delms$ and $\delmd$ would provide a powerful constraint
on the CKM matrix elements. The ratio of these mass differences can be
expressed in the SM as:
\beq
\frac{\delms}{\delmd} =
 \frac{\hat{\eta}_{B_s}M_{B_s}\left(\fbbs\right)}
{\hat{\eta}_{B_d}M_{B_d}\left(\fbb\right)}
\left\vert \frac{V_{ts}}{V_{td}} \right\vert^2
=C\frac{\xi_s^2}{\lambda^2} \frac{1}{(1-\bar{\rho})^2 + \bar{\eta}^2}.
\label{xratio}
\eeq
Since the QCD correction factors satisfy $\hat{\eta}_{B_s} =
\hat{\eta}_{B_d} = 0.55$ \cite{etaB}, and since $C = M_{B_s}/M_{B_d} =
1.017$ \cite{PDG00}, the only real uncertainty in this quantity is the
ratio of hadronic matrix elements $\xi_s \equiv (f_{B_s}
\sqrt{\hat{B}_{B_s}}) / (f_{B_d}\sqrt{\hat{B}_{B_d}})$. It is now
widely accepted that the ratio $\xi_s$ is probably the most reliable
of the lattice-QCD estimates in $B$ physics, $\xi_s=1.16 \pm 0.05$
\cite{Bernard00}. Thus, the accurate knowledge of $\Delta M_s/\Delta
M_d$ puts a stringent constraint on the CKM parameters $\bar{\rho}$
and $\bar{\eta}$, and hence on the allowed region of the unitarity
triangle.

Since $\delmd$ has already been measured very accurately (the present
world average is $\delmd=0.487 \pm 0.014 ~{\rm ps}^{-1}$
\cite{Stocchi00}), a measurement of $\delms$ is being keenly awaited.
The present experimental situation on $\delms$ can be summarized as
follows: the combined analysis of the LEP/SLD/CDF measurements
undertaken by the $B$-oscillation working group yields a lower bound
$\delms > 14.9 ~{\rm ps}^{-1}$ (at 95\% C.L.) \cite{Stocchi00}, using
the amplitude analysis method of Moser and Rousarie \cite{Moser97}.
However, quite interestingly, the same analysis also yields a {\it
  local minimum in the log-likelihood distribution around} $\delms
=17.7 ~{\rm ps}^{-1}$, whose significance becomes more pronounced if
the amplitude spectrum is converted to a log-likelihood function
referenced to $\delms =\infty$: $\Delta\log {\it L}^\infty(\delms) =
(0.5 -{\cal A})/\sigma_{\cal A}^2$ \cite{checciaetal}. Here ${\cal A}$
is an amplitude modulating the oscillating terms as $ (1\pm {\cal A}
\cos \delms t)$, with $\sigma_{\cal A}$ being its error. This local
minimum has the interpretation that at this value of $\delms$, the
amplitude ${\cal A}$ is away from being zero (no-mixing case) by
$2.5\sigma$. The statistical significance of this result has been
studied in a monte-carlo based analysis by Boix and Abbaneo
\cite{boix-abbaneo}. They estimate the probability that the observed
result was produced by a statistical fluctuation anywhere in the
scanned values of $\delms$ to be $1 - {\rm C.L.} \simeq 2.5\%$
\cite{Stocchi00}. Although this probability is not yet small enough to
consider this to be a measurement of $\Delta M_s$, the result is
intriguing.

The other quantity which must be mentioned is $\sin 2 \beta$. Since a
non-zero value of $\sin 2\beta$ would be the first evidence for CP
violation outside the kaon system, many experiments are attempting to
measure this quantity. In the Wolfenstein parametrization, $-\beta$ is
the phase of the CKM matrix element $V_{td}$. From Eq.~(\ref{CKM}) one
can readily find that
\beq
\sin 2 \beta = \frac{2\bar{\eta}(1-\bar{\rho})}{(1-\bar{\rho})^2 +   
\bar{\eta}^2} ~.
\label{sin2beta-wolf}
\eeq
Thus, a measurement of $\sin 2\beta$ would put a strong constraint on
the parameters $\bar{\rho}$ and $\bar{\eta}$.

In fact, first measurements of $\sin 2 \beta$ have already been
reported, and the present status is summarized below:
\begin{eqnarray}
\sin 2 \beta &=& 3.2 ^{+1.8}_{-2.0} \pm 0.5
~~(\mbox{OPAL \cite{OPAL-sin2beta}}),\\ \nonumber
&=& 0.79 ^{+0.41}_{-0.44}~~(\mbox{CDF \cite{CDF-sin2beta}}),\\ \nonumber
& =& 0.84 ^{+0.83}_{-1.04} \pm 0.16 ~~(\mbox{ALEPH
\cite{ALEPH-sin2beta}}),\\ \nonumber
& =& 0.45^{+0.43 ~+0.07}_{-0.44 ~-0.09}  ~~(\mbox{BELLE
\cite{BELLE-sin2beta}}), \\ \nonumber
& =& 0.12 \pm 0.37 \pm 0.09~~(\mbox{BABAR \cite{BABAR-sin2beta}}),
\label{sin2beta-all}
\end{eqnarray}
yielding a world average $\sin 2 \beta = 0.48^{+0.22}_{-0.24}$
\cite{Faccioli}. This quantity will eventually be very precisely
measured at the ongoing $B$-factory experiments and elsewhere.
However, since the error is still quite large, we do not include this
measurement in our fits.

\subsection{SM Fits}

\begin{figure}
\vskip -1.0truein
\centerline{\epsfxsize 3.5 truein \epsfbox {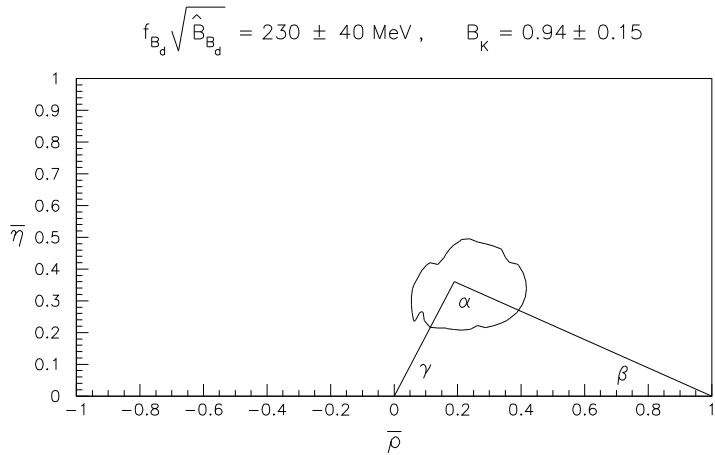}}
\vskip -1.5truein
\caption{Allowed region in $\bar\rho$--$\bar\eta$ space in the SM,
from a fit to the ten parameters discussed in the text and given in
Table \protect{\ref{datatable}}. The solid line represents the region
with $\chi^2=\chi_{min}^2+6$ corresponding to the 95\% C.L.\
region. The triangle shows the best fit.}
\label{rhoeta1}
\end{figure}

In order to find the allowed region in $\bar{\rho}$--$\bar{\eta}$
space, i.e.\ the allowed shapes of the unitarity triangle, the
computer program MINUIT is used to fit the parameters to all the
experimental constraints. In the fit, we allow ten parameters to vary:
$\bar{\rho}$, $\bar{\eta}$, $A$, $m_t$, $m_c$, $\eta_{cc}$,
$\eta_{ct}$, $f_{B_d} \sqrt{\hat{B}_{B_d}}$, $\hat{B}_K$, and $\xi_s$.
The $\Delta M_s$ constraint is included using the amplitude method
\cite{Moser97}. The allowed (95\% C.L.) $\bar\rho$--$\bar\eta$ region
is shown in Fig.~\ref{rhoeta1}. The triangle drawn is to facilitate
our discussions, and corresponds to the central values of the fits,
$(\alpha,\beta,\gamma) = (95^\circ,22^\circ,63^\circ)$.

The CP angles $\alpha$, $\beta$ and $\gamma$ can be measured in
CP-violating rate asymmetries in $B$ decays.  These angles can be
expressed in terms of $\bar{\rho}$ and $\bar{\eta}$. Thus, different
shapes of the unitarity triangle are equivalent to different values of
the CP angles. Referring to Fig.~\ref{rhoeta1}, the allowed ranges at
95\% C.L. are given by
\beq
77^\circ \le \alpha \le 127^\circ ~~,~~~~
14^\circ \le \beta \le 35^\circ ~~,~~~~
34^\circ \le \gamma \le 81^\circ ~~,
\label{CPangleregion}
\eeq
or, equivalently,
\beq
-0.96 \le  \sin 2\alpha  \le 0.45 ~~,~~~~
0.46  \le  \sin 2\beta  \le 0.94  ~~,~~~~
0.31  \le  \sin^2 \gamma  \le 0.98 ~~.
\label{smabgrange}
\eeq

\begin{figure}
\vskip -2.0truein
\centerline{\epsfxsize 6.0 truein \epsfbox {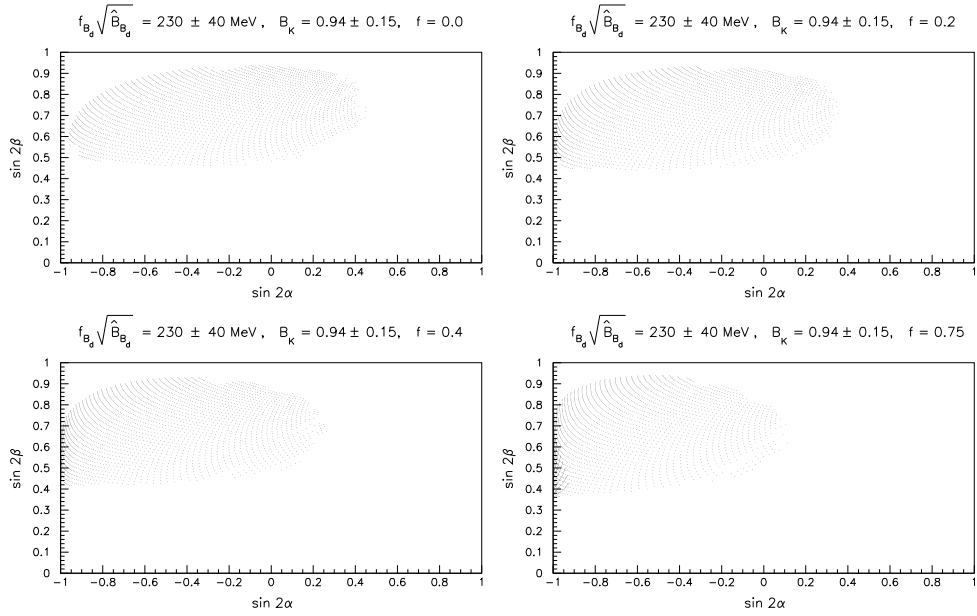}}
\vskip -3.7truein
\caption{Allowed 95\% C.L. region of the CP-violating quantities 
  $\sin 2\alpha$ and $\sin 2\beta$, from a fit to the data given in
  Table \protect{\ref{datatable}}. The upper left plot ($f=0$)
  corresponds to the SM, while the other plots ($f=0.2$, 0.4, 0.75)
  correspond to various SUSY models.}
\label{alphabetacorrsm}
\end{figure}

\begin{figure}
\vskip -2.0truein
\centerline{\epsfxsize 6.0 truein \epsfbox {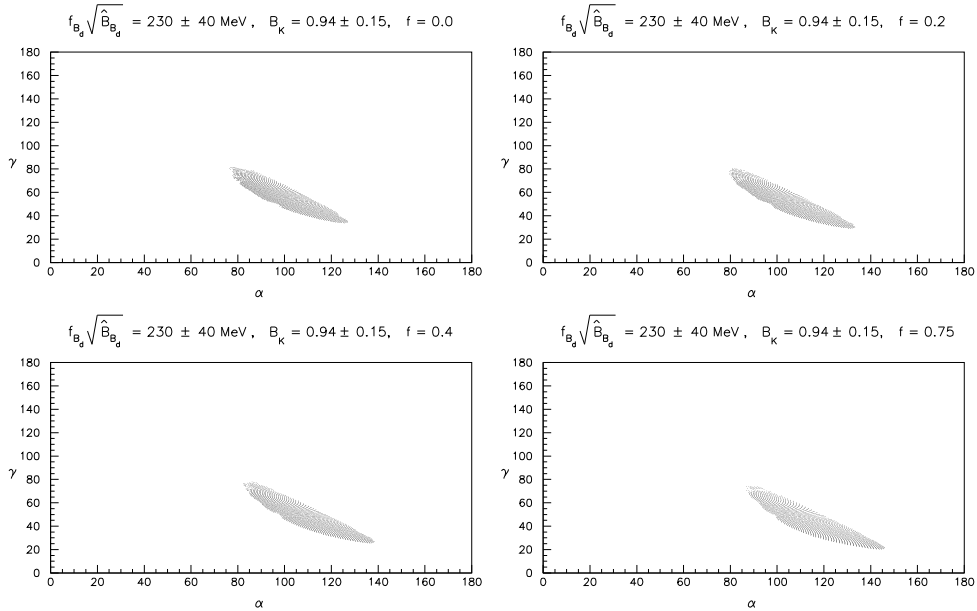}}
\vskip -3.7truein
\caption{Allowed 95\% C.L. region of the CP-violating quantities 
  $\alpha$ and $\gamma$, from a fit to the data given in Table
  \protect{\ref{datatable}}. The upper left plot ($f=0$) corresponds
  to the SM, while the other plots ($f=0.2$, 0.4, 0.75) correspond to
  various SUSY models.}
\label{alphagammacorrsm}
\end{figure}

Of course, the values of $\alpha$, $\beta$ and $\gamma$ are
correlated, i.e.\ they are not all allowed simultaneously. After all,
the sum of these angles must equal $180^\circ$. We illustrate these
correlations in Figs.~\ref{alphabetacorrsm} and
\ref{alphagammacorrsm}. In both of these figures, the SM plot is
labelled by $f=0$. Fig.~\ref{alphabetacorrsm} shows the allowed region
in $\sin 2\alpha$--$\sin 2\beta$ space allowed by the data, while
Fig.~\ref{alphagammacorrsm} shows the allowed (correlated) values of
the CP angles $\alpha$ and $\gamma$. This correlation is roughly
linear, due to the relatively small allowed range of $\beta$
[Eq.~(\ref{CPangleregion})].

Finally, one can also calculate the range of $\Delta M_s$ which is
presently allowed in the SM. At 95\% C.L. we find:
\beq
14.6 \le \Delta M_s \le 31.2 ~.
\eeq

\subsection{SUSY Fits}

In this subsection we update the profile of the unitarity triangle in
supersymmetric (SUSY) theories with minimal flavor violation. In this
class of models, the SUSY contributions to $\delmd$, $\delms$ and
$\abseps$ can all be described by a single common parameter $f$ (for a
more detailed discussion, we refer the reader to Ref.~\cite{AL99}):
\begin{eqnarray}
\delmd &=& \delmd (SM) [ 1 + f ], \nonumber \\
\delms &=& \delms (SM) [ 1 + f ], \nonumber \\
\abseps &=& \frac{G_F^2f_K^2M_KM_W^2}{6\sqrt{2}\pi^2\Delta M_K}
\hat{B}_K\left(A^2\lambda^6\bar{\eta}\right)
\bigl(y_c\left\{\hat{\eta}_{ct}f_3(y_c,y_t)-\hat{\eta}_{cc}\right\}
 \nonumber \\
&~& + ~~~~~~~~~
\hat{\eta}_{tt}y_tf_2(y_t)[1 + f] A^2\lambda^4(1-\bar{\rho})\bigr).
\label{susyformel}
\end{eqnarray}
The parameter $f$ is positive definite, so that the supersymmetric
contributions add {\it constructively} to the SM contributions in the
entire allowed supersymmetric parameter space. The size of $f$
depends, in general, on the parameters of the supersymmetric model.
In our fits, we will consider four representative values of $f$ --- 0,
0.2, 0.4 and 0.75 --- which are typical of the SM, minimal
supergravity (SUGRA) models, non-minimal SUGRA models, and non-SUGRA
models, respectively.

\begin{figure}
\vskip -1.0truein
\centerline{\epsfxsize 3.5 truein \epsfbox {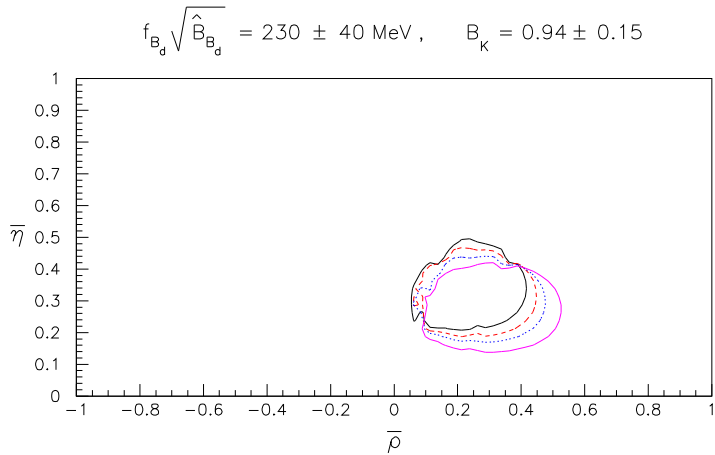}}
\vskip -1.5truein
\caption{Allowed 95\% C.L. region in $\rho$--$\eta$ space in the SM 
  and in SUSY models, from a fit to the data given in Table
  \protect{\ref{datatable}}. From left to right, the allowed regions
  correspond to $f=0$ (SM, solid line), $f=0.2$ (long dashed line),
  $f=0.4$ (short dashed line), $f=0.75$ (dotted line).}
\label{sugratot}
\end{figure}

For the SUSY fits, we use the same program as for the SM fits, except
that the theoretical expressions for $\Delta M_d$, $\Delta M_s$ and
$\abseps$ are modified as above [Eq.~(\ref{susyformel})]. The allowed
95\% C.L. regions for the four values $f=0$, 0.2, 0.4, and 0.75 are
all plotted in Fig.~\ref{sugratot}. We can see from this figure that,
as $f$ increases, the allowed region moves slightly down and towards
the right in the $\bar{\rho}$--$\bar{\eta}$ plane.

At present, there is still considerable overlap between the $f=0$ (SM)
and $f=0.75$ regions.  However, there are also regions allowed for one
value of $f$ which are excluded for another value. In particular, one
notices that, as $f$ increases, larger values of ${\bar\rho}$ are
allowed. This in turn implies that larger values of $\Delta M_s$ are
allowed, and in fact this is borne out quantitively.  The allowed
ranges for $\Delta M_s$ (95\% C.L.) are given by:
\begin{eqnarray}
f = 0 ~{\rm (SM)} & : & 14.6 \le \Delta M_s \le 31.2 ~, \nn \\
f = 0.2 & : & 14.6 \le \Delta M_s \le 35.5 ~, \nn \\
f = 0.4 & : & 14.9 \le \Delta M_s \le 39.4 ~, \nn \\
f = 0.75 & : & 15.1 \le \Delta M_s \le 48.6 ~. 
\end{eqnarray}
Although the lower limit on $\Delta M_s$ is roughly independent of
$f$, the upper limit increases as $f$ increases. Thus, should $\Delta
M_s$ be found to be very large, this would be consistent with SUSY
models with large values of $f$. Conversely, if $\Delta M_s$ is
measured to be near its lower limit, this would disfavor SUSY models
with large $f$. (Note that, although small values of $\Delta M_s$ are
allowed in such models, the region of parameter space which yields
such values is relatively small. Thus, one can expect the fits to the
data to be poorer for SUSY models with large values of $f$ than for
models with small $f$. We will see this in more detail in Sec.~3.)

As was seen in the SM fit, different shapes of the unitarity triangle
correspond to different values of the CP phases $\alpha$, $\beta$ and
$\gamma$. Furthermore, these allowed values are correlated: the
correlations between $\sin 2\alpha$ and $\sin 2\beta$, and between
$\alpha$ and $\gamma$, are shown in Fig.~\ref{alphabetacorrsm} and
Fig.~\ref{alphagammacorrsm}, respectively. Tables~\ref{cpasym1} and
\ref{cpasym2} give, respectively, the allowed ranges for the CP phases
and the quantities measured in CP-violating asymmetries. The key
observation here is that a measurement of the CP angle $\beta$ will
{\it not} distinguish among the various values of $f$ -- the allowed
range for $\beta$ is rather independent of $f$. If one wants to
distinguish among the various SUSY models, it will be necessary to
measure $\alpha$ and/or $\gamma$. (Of course, as mentioned above,
there is still significant overlap among all four models. Thus,
depending on what values of $\alpha$ and $\gamma$ are obtained, we may
or may not be able to rule out certain values of $f$.)

\begin{table}
\hfil
\caption{Allowed 95\% C.L. ranges for the CP phases $\alpha$, $\beta$
and $\gamma$, as well as their central values, from the CKM fits in
the SM $(f=0)$ and supersymmetric theories, characterized by the
parameter $f$ defined in the text.}
\vbox{\offinterlineskip
\halign{&\vrule#&
 \strut\quad#\hfil\quad\cr
\noalign{\hrule}
height2pt&\omit&&\omit&&\omit&&\omit&&\omit&\cr 
& $f$ && $\alpha$ && $\beta$ && $\gamma$ && $(\alpha,\beta,\gamma)_{\rm cent}$ & \cr
height2pt&\omit&&\omit&&\omit&&\omit&&\omit&\cr 
\noalign{\hrule}
height2pt&\omit&&\omit&&\omit&&\omit&&\omit&\cr
& $f=0$ (SM) && $77^\circ$ -- $127^\circ$ && $14^\circ$ -- $35^\circ$ &&
$34^\circ$ -- $81^\circ$ && $(95^\circ, 22^\circ, 63^\circ)$ & \cr
& $f=0.2$ && $80^\circ$ -- $133^\circ$ && $13^\circ$ -- $34^\circ$ &&
$29^\circ$ -- $81^\circ$ && $(109^\circ, 22^\circ, 49^\circ)$ & \cr   
& $f=0.4$ && $82^\circ$ -- $138^\circ$ && $12^\circ$ -- $34^\circ$ &&
$25^\circ$ -- $78^\circ$ && $(112^\circ, 20^\circ, 48^\circ)$ & \cr
& $f=0.75$ && $87^\circ$ -- $146^\circ$ && $10^\circ$ -- $35^\circ$ &&
$20^\circ$ -- $74^\circ$ && $(114^\circ, 21^\circ, 45^\circ)$ & \cr
height2pt&\omit&&\omit&&\omit&&\omit&&\omit&\cr
\noalign{\hrule}}}
\label{cpasym1}
\end{table}

\begin{table}
\caption{Allowed 95\% C.L. ranges for the CP asymmetries $\sin
2\alpha$, $\sin 2\beta$ and $\sin^2 \gamma$, from the CKM fits in the
SM $(f=0)$ and supersymmetric theories, characterized by the parameter
$f$ defined in the text.}  
\hfil \vbox{\offinterlineskip
\halign{&\vrule#& \strut\quad#\hfil\quad\cr \noalign{\hrule}
height2pt&\omit&&\omit&&\omit&&\omit&\cr & $f$ && $\sin 2\alpha$ &&
$\sin 2\beta$ && $\sin^2 \gamma$ & \cr
height2pt&\omit&&\omit&&\omit&&\omit&\cr \noalign{\hrule}
height2pt&\omit&&\omit&&\omit&&\omit&\cr
& $f=0$ (SM) && $-$0.96 -- 0.45 && 0.46 -- 0.94 && 0.31 -- 0.98 & \cr
& $f=0.2$ && $-$1.00 -- 0.35 && 0.44 -- 0.93 && 0.24 -- 0.97 & \cr
& $f=0.4$ && $-$1.00 -- 0.26 && 0.42 -- 0.93 && 0.19 -- 0.96 & \cr
& $f=0.75$ && $-$1.00 -- 0.11 && 0.36 -- 0.94 && 0.12 -- 0.93 & \cr
height2pt&\omit&&\omit&&\omit&&\omit&\cr \noalign{\hrule}}}
\label{cpasym2}
\end{table}

\section{Unitarity Triangle: Future Profile}

As was discussed in Sec.~2.1, the $\bs$--$\bsbar$ mixing data appears
to contain a $2.5\sigma$ signal centered at $\Delta M_s = 17.7~{\rm
  ps}^{-1}$. This signal is not statistically significant enough to be
considered a measurement of $\Delta M_s$. However, it is still
interesting to consider what the effect would be on the profile of the
unitarity triangle, both in the SM and in SUSY models, if this signal
persisted and became a measurement. This is the purpose of this section.

In order to be consistent with both the central value of $\Delta M_s$
and its 95\% C.L. lower limit ($14.9~{\rm ps}^{-1}$), we assume the
hypothetical future measurement of this quantity to be
\beq
\Delta M_s = 17.7 \pm 1.4 ~{\rm ps}^{-1} ~.
\label{Dmsmeas}
\eeq
The SM and SUSY fits are then performed with this as part of the input
data.

\begin{figure}
\vskip -1.0truein
\centerline{\epsfxsize 3.5 truein \epsfbox {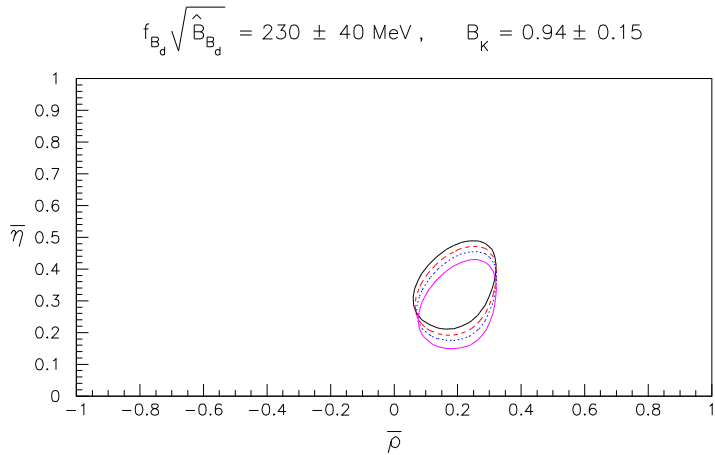}}
\vskip -1.5truein
\caption{Allowed 95\% C.L. region in $\rho$--$\eta$ space in the SM 
  and in SUSY models, in the hypothetical scenario in which $\Delta
  M_s$ is given by Eq.~(\protect\ref{Dmsmeas}). From top to bottom,
  the allowed regions correspond to $f=0$ (SM, solid line), $f=0.2$
  (long dashed line), $f=0.4$ (short dashed line), $f=0.75$ (dotted
  line).}
\label{dmssugratot}
\end{figure}

The results are shown in Fig.~\ref{dmssugratot}. The effect of the
$\Delta M_s$ constraint is quite striking: the minimum- and
maximum-allowed values of ${\bar\rho}$ are essentially independent of
$f$. Now, as $f$ increases, the allowed region only moves slightly
down in the $\bar{\rho}$--$\bar{\eta}$ plane.

\begin{figure}
\vskip -1.0truein
\centerline{\epsfxsize 3.0 truein \epsfbox {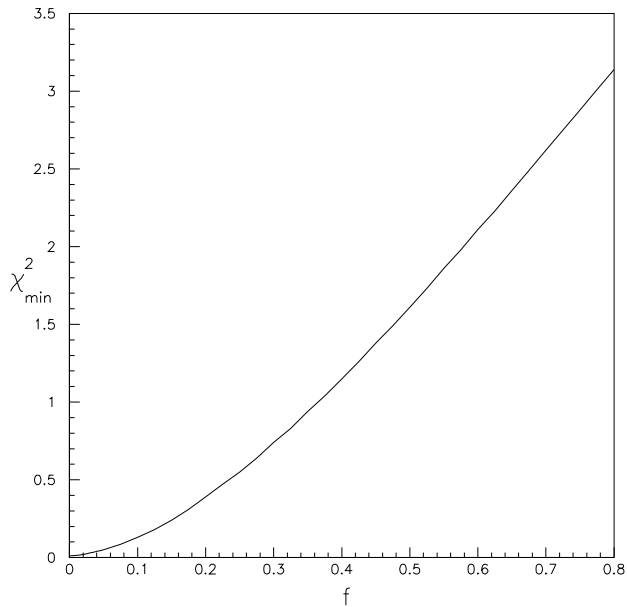}}
\vskip -0.5truein
\caption{Minimum value of $\chi^2$ as a function of the SUSY parameter
  $f$, for the fits in the hypothetical scenario in which $\Delta M_s$
  is given by Eq.~(\protect\ref{Dmsmeas}). Models with $\chi^2_{min} >
  2$ are disfavored.}
\label{fchi2min}
\end{figure}

However, Fig.~\ref{dmssugratot} does not tell the whole story. In
particular, it does not take into account how good the fits are for
the various values of $f$. The goodness of fit is indicated by the
minimum value of $\chi^2$: since there are two degrees of freedom
($\bar{\rho}$ and $\bar{\eta}$), fits with $\chi^2_{min} > 2$ are
disfavored. In fact, the model with $f=0.75$ has $\chi^2_{min} =
2.9$, and is hence a poor fit to the data.  In Fig.~\ref{fchi2min} we
present $\chi^2_{min}$ as a function of $f$. This figure shows that,
for the hypothetical scenario in which $\Delta M_s$ given by
Eq.~(\ref{Dmsmeas}), models with $f > 0.6$ are disfavored.

It is interesting --- and perhaps somewhat discouraging --- to note
that the best fit ($\chi^2_{min} = 9.5 \times 10^{-3}$) occurs for
$f=0$, i.e.\
for the standard model.
That is, although some models with $f \ne 0$ would give reasonable
fits to the data, the hint of a signal at $\Delta M_s = 17.7~{\rm
ps}^{-1}$ does not indicate any problems whatsoever for the SM.

Note also that the percentage error we have assumed for $\Delta M_s$,
7.9\%, is considerably greater than the present experimental error on
$\Delta M_d$ of 2.9\%. It is not unreasonable to believe that the
percentage error on $\Delta M_s$ will eventually approach that of
$\Delta M_d$. In that case, the precise measurement of $\Delta M_s$
will be able to rule out an even greater region of SUSY parameter
space. That is, values of $f$ smaller than 0.6 will be disfavored.
Thus, we see that a precision measurement of $\Delta M_s$ will be an
extremely powerful tool for distinguishing among the SM and its
various supersymmetric extensions.

\begin{figure}
\vskip -2.0truein
\centerline{\epsfxsize 6.0 truein \epsfbox {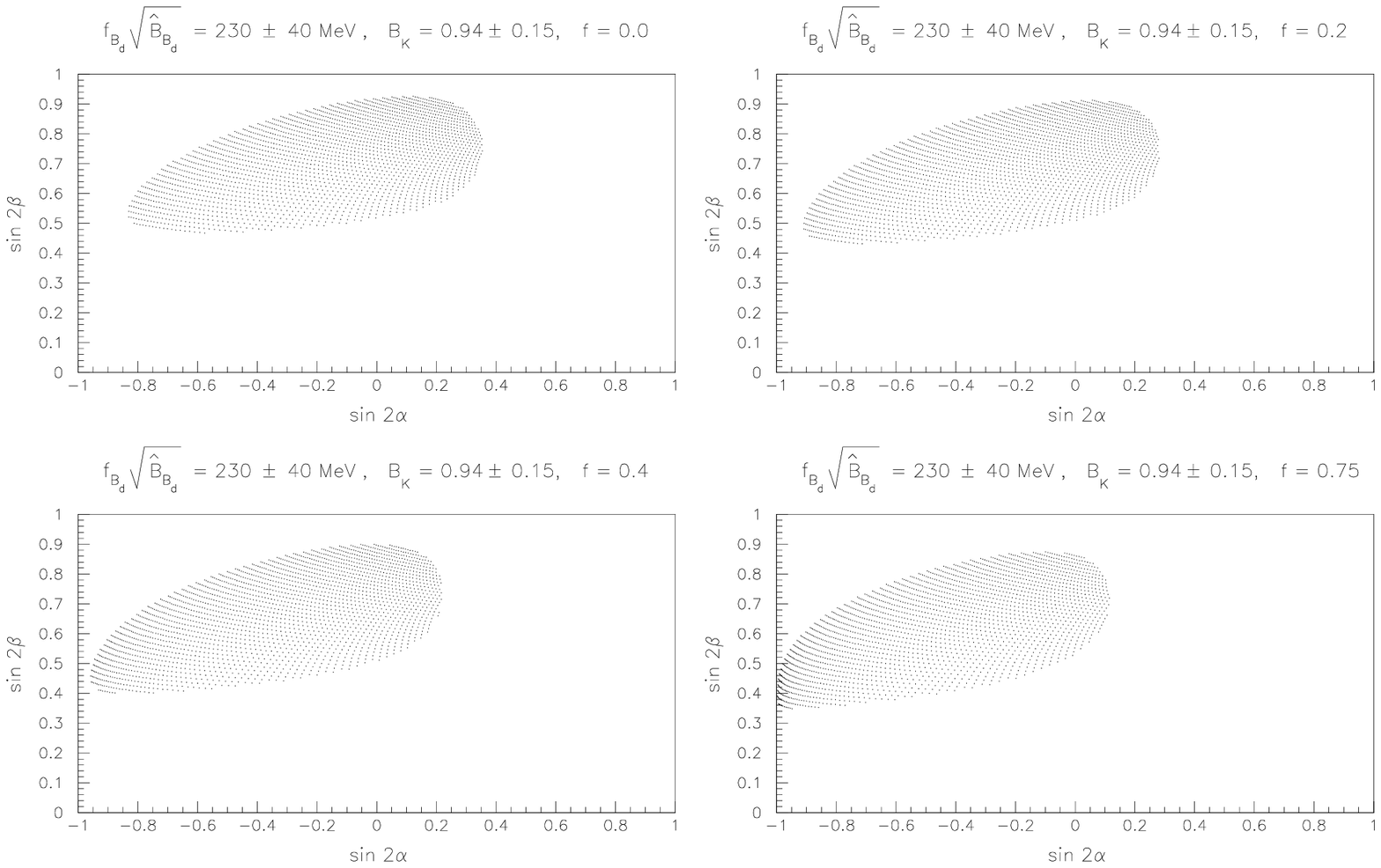}}
\vskip -3.7truein
\caption{Allowed 95\% C.L. region of the CP-violating quantities $\sin
  2\alpha$ and $\sin 2\beta$, in the hypothetical scenario in which
  $\Delta M_s$ is given by Eq.~(\protect\ref{Dmsmeas}). The upper left
  plot ($f=0$) corresponds to the SM, while the other plots ($f=0.2$,
  0.4, 0.75) correspond to various SUSY models.}
\label{dmsalphabetacorrsm}
\end{figure}

\begin{figure}
\vskip -2.0truein
\centerline{\epsfxsize 6.0 truein \epsfbox {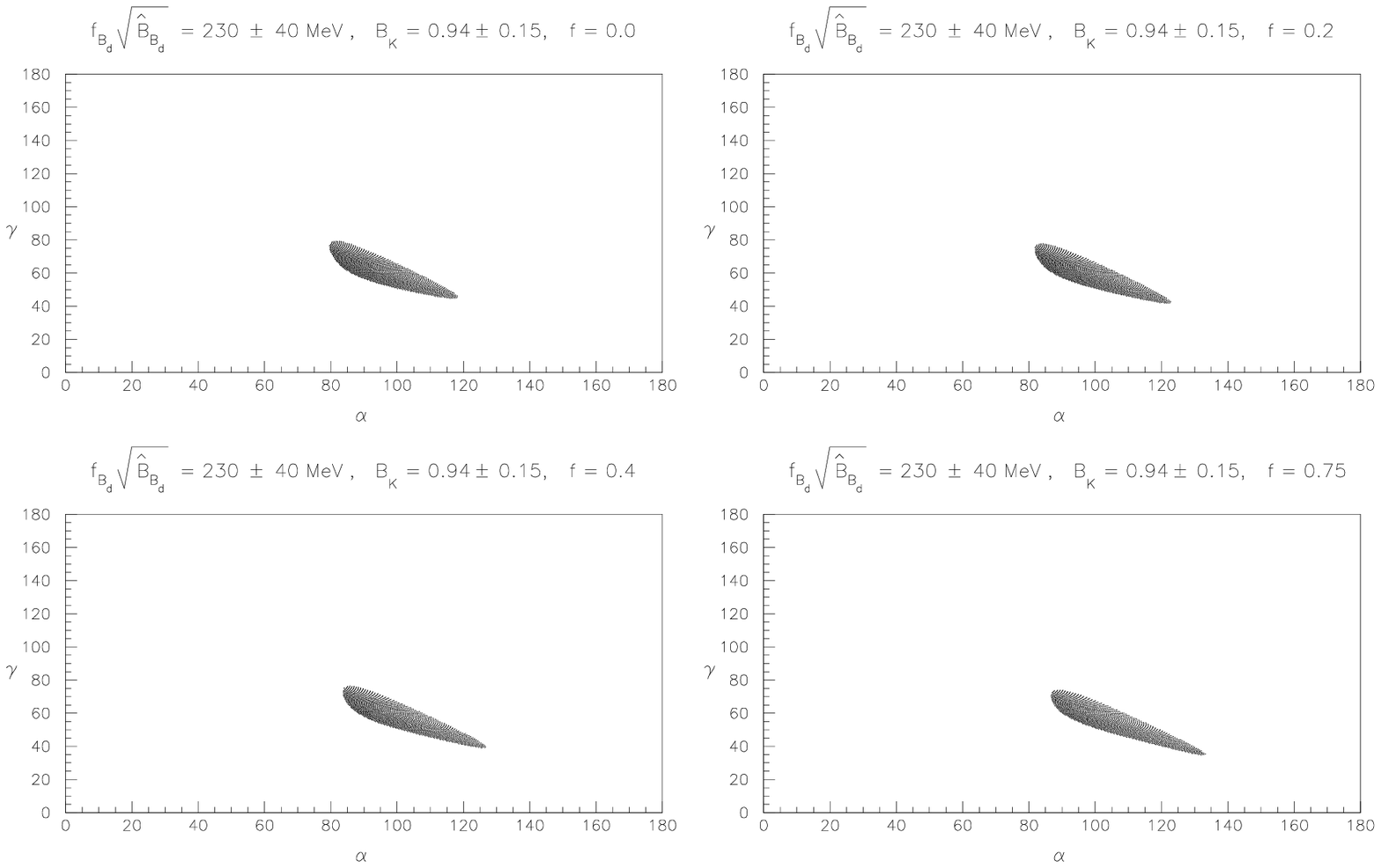}}
\vskip -3.7truein
\caption{Allowed 95\% C.L. region of the CP-violating quantities
  $\alpha$ and $\gamma$, in the hypothetical scenario in which
  $\Delta M_s$ is given by Eq.~(\protect\ref{Dmsmeas}). The upper left
  plot ($f=0$) corresponds to the SM, while the other plots ($f=0.2$,
  0.4, 0.75) correspond to various SUSY models.}
\label{dmsalphagammacorrsm}
\end{figure}

For completeness, in Figs.~\ref{dmsalphabetacorrsm} and
\ref{dmsalphagammacorrsm} we present, respectively, the $\sin
2\alpha$--$\sin 2\beta$ and $\alpha$--$\gamma$ correlations for the
scenario in which $\Delta M_s$ is given by Eq.~(\ref{Dmsmeas}). The
allowed ranges for the CP phases and for $\sin 2\alpha$, $\sin 2\beta$
and $\sin^2 \gamma$ are given in Tables~\ref{cpasym3} and
\ref{cpasym4}, respectively. A comparison of, for example, Tables
\ref{cpasym1} and \ref{cpasym3} reveals that, as expected, the
measurement of $\Delta M_s$ does not affect the allowed range for
$\beta$ appreciably, though the ranges for $\alpha$ and $\gamma$ are
significantly reduced.

\begin{table}
\hfil
\caption{Allowed 95\% C.L. ranges for the CP phases $\alpha$, $\beta$ 
  and $\gamma$, as well as their central values, from the CKM fits in
  the SM $(f=0)$ and supersymmetric theories, in the hypothetical 
  scenario in which $\Delta M_s$ is given by Eq.~(\protect\ref{Dmsmeas}).}
\vbox{\offinterlineskip
\halign{&\vrule#&
 \strut\quad#\hfil\quad\cr
\noalign{\hrule}
height2pt&\omit&&\omit&&\omit&&\omit&&\omit&\cr
& $f$ && $\alpha$ && $\beta$ && $\gamma$ && $(\alpha,\beta,\gamma)_{\rm
cent}$ & \cr 
height2pt&\omit&&\omit&&\omit&&\omit&&\omit&\cr
\noalign{\hrule} 
height2pt&\omit&&\omit&&\omit&&\omit&&\omit&\cr
& $f=0$ (SM) && $80^\circ$ -- $119^\circ$ && $14^\circ$ -- $34^\circ$ &&
$44^\circ$ -- $79^\circ$ && $(98^\circ, 22^\circ, 60^\circ)$ & \cr
& $f=0.2$ && $82^\circ$ -- $123^\circ$ && $13^\circ$ -- $33^\circ$ &&
$41^\circ$ -- $77^\circ$ && $(101^\circ, 21^\circ, 58^\circ)$ & \cr   
& $f=0.4$ && $84^\circ$ -- $127^\circ$ && $12^\circ$ -- $32^\circ$ &&
$38^\circ$ -- $76^\circ$ && $(105^\circ, 20^\circ, 55^\circ)$ & \cr
& $f=0.75$ && $87^\circ$ -- $134^\circ$ && $10^\circ$ -- $30^\circ$ &&
$34^\circ$ -- $73^\circ$ && $(110^\circ, 18^\circ, 52^\circ)$ & \cr
height2pt&\omit&&\omit&&\omit&&\omit&&\omit&\cr
\noalign{\hrule}}}
\label{cpasym3}
\end{table}

\begin{table}
\caption{Allowed 95\% C.L. ranges for the CP asymmetries $\sin
  2\alpha$, $\sin 2\beta$ and $\sin^2 \gamma$, from the CKM fits in the
  SM $(f=0)$ and supersymmetric theories, in the hypothetical 
  scenario in which $\Delta M_s$ is given by Eq.~(\protect\ref{Dmsmeas}).}
\hfil \vbox{\offinterlineskip
\halign{&\vrule#& \strut\quad#\hfil\quad\cr \noalign{\hrule}
height2pt&\omit&&\omit&&\omit&&\omit&\cr & $f$ && $\sin 2\alpha$ &&
$\sin 2\beta$ && $\sin^2 \gamma$ & \cr
height2pt&\omit&&\omit&&\omit&&\omit&\cr \noalign{\hrule}
height2pt&\omit&&\omit&&\omit&&\omit&\cr
& $f=0$ (SM) && $-$0.84 -- 0.35 && 0.47 -- 0.93 && 0.48 -- 0.96 & \cr
& $f=0.2$ && $-$0.92 -- 0.27 && 0.44 -- 0.91 && 0.43 -- 0.95 & \cr
& $f=0.4$ && $-$0.97 -- 0.20 && 0.40 -- 0.90 && 0.38 -- 0.94 & \cr
& $f=0.75$ && $-$1.00 -- 0.09 && 0.35 -- 0.87 && 0.31 -- 0.92 & \cr
height2pt&\omit&&\omit&&\omit&&\omit&\cr \noalign{\hrule}}}
\label{cpasym4} 
\end{table}

\section{Conclusions}

The latest experimental data on $\bs$--$\bsbar$ mixing puts the 95\%
C.L. lower limit at $\delms > 14.9 ~{\rm ps}^{-1}$. Furthermore, there
is an intriguing $2.5\sigma$ hint of a signal at $\delms \simeq
17.7~{\rm ps}^{-1}$. In light of this, in this paper we examine the
effect that a measurement of $\Delta M_s$ would have on the profile of
the CKM matrix, both in the standard model and in supersymmetric
models with minimal flavor violation.

We first update the profile of the unitarity triangle, both in the SM
and in supersymmetric models, using current experimental data. The
SUSY contributions to $\delmd$, $\delms$ and $\abseps$ can all be
described by a single common parameter $f$, and we take three
representative values in our fits: $f = 0.2$, 0.4 and 0.75. The
measurement of the CP-phase $\beta$ will not distinguish among the
various models, though the measurement of $\alpha$ and/or $\gamma$ may
do so. More importantly, the different models make different
predictions for the allowed range of $\Delta M_s$. This indicates that
the measurement of $\Delta M_s$ will also be important for
distinguishing among the various models.

This point is made quantitatively when the fits are repeated assuming
a hypothetical measurement of $\Delta M_s = 17.7 \pm 1.4~{\rm
  ps}^{-1}$. In this case, we find that SUSY models with $f > 0.6$
provide poor fits to the data, and are hence disfavored. Thus, we see
that the measurement of $\delms$ is indeed a powerful tool for
discriminating between the SM and its supersymmetric extensions.

For the particular experimental value of $\delms$ that we have assumed
--- and we have chosen this value to be consistent with the lower 95\%
C.L. bound, as well as with the hint of a signal --- the best fit to
the data occurs for $f=0$, i.e.\ for the SM. Thus, present data on
$\bs$--$\bsbar$ mixing does not indicate any problems with the SM,
which may be somewhat discouraging for those who hope to see signals
for new physics via CKM phenomenology.

Finally, it is not unreasonable to expect that the percentage error on
a measurement of $\Delta M_s$ will eventually reach the same level as
that of $\Delta M_d$ (i.e.\ $\sim 3$\%). When this happens, the
precise measurement of $\Delta M_s$ will be able to rule out an even
greater region of SUSY parameter space. (Or, if the central value
changes, one could conceivably rule out the SM!) Once again, this
emphasizes the importance of a measurement of $\delms$ for searching
for new flavor physics.

\bigskip
\noindent
{\bf Acknowledgements}:
\bigskip
We would like to thank John Jaros, Fabrizio Parodi, and Achille
Stocchi for very helpful discussions about the current measurements
and fit procedures used in constraining $\delms$. The work of D.L.
was financially supported by NSERC of Canada.
  
\newpage


\end{document}